\documentclass[a4paper,11pt]{article}
\usepackage{amssymb}
\usepackage{amsbsy}
\usepackage{amsmath}
\usepackage{graphicx}
\usepackage[latin1]{inputenc} 
\usepackage{atxy}
\usepackage{fancybox}
\usepackage{color}
\usepackage[T1]{fontenc}

\begin{document}

\titlepage{
\atxy(0.5cm,0.5cm){
\includegraphics[width=7.0cm]{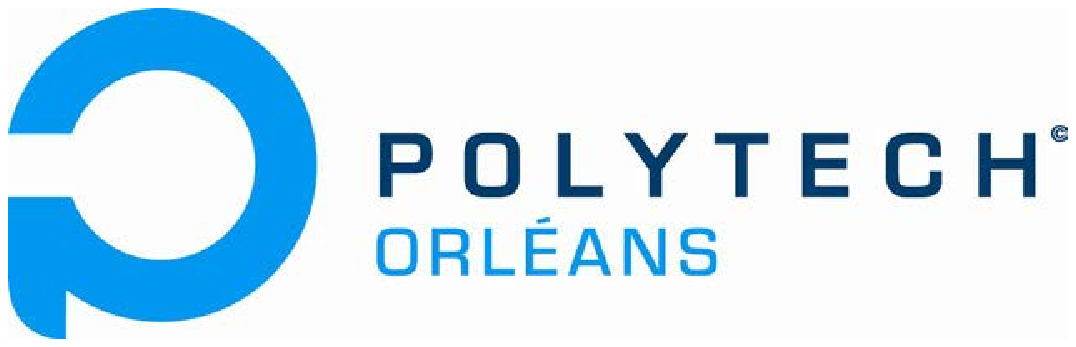}}

\atxy(16.0cm,0.3cm){
\includegraphics[width=2.5cm]{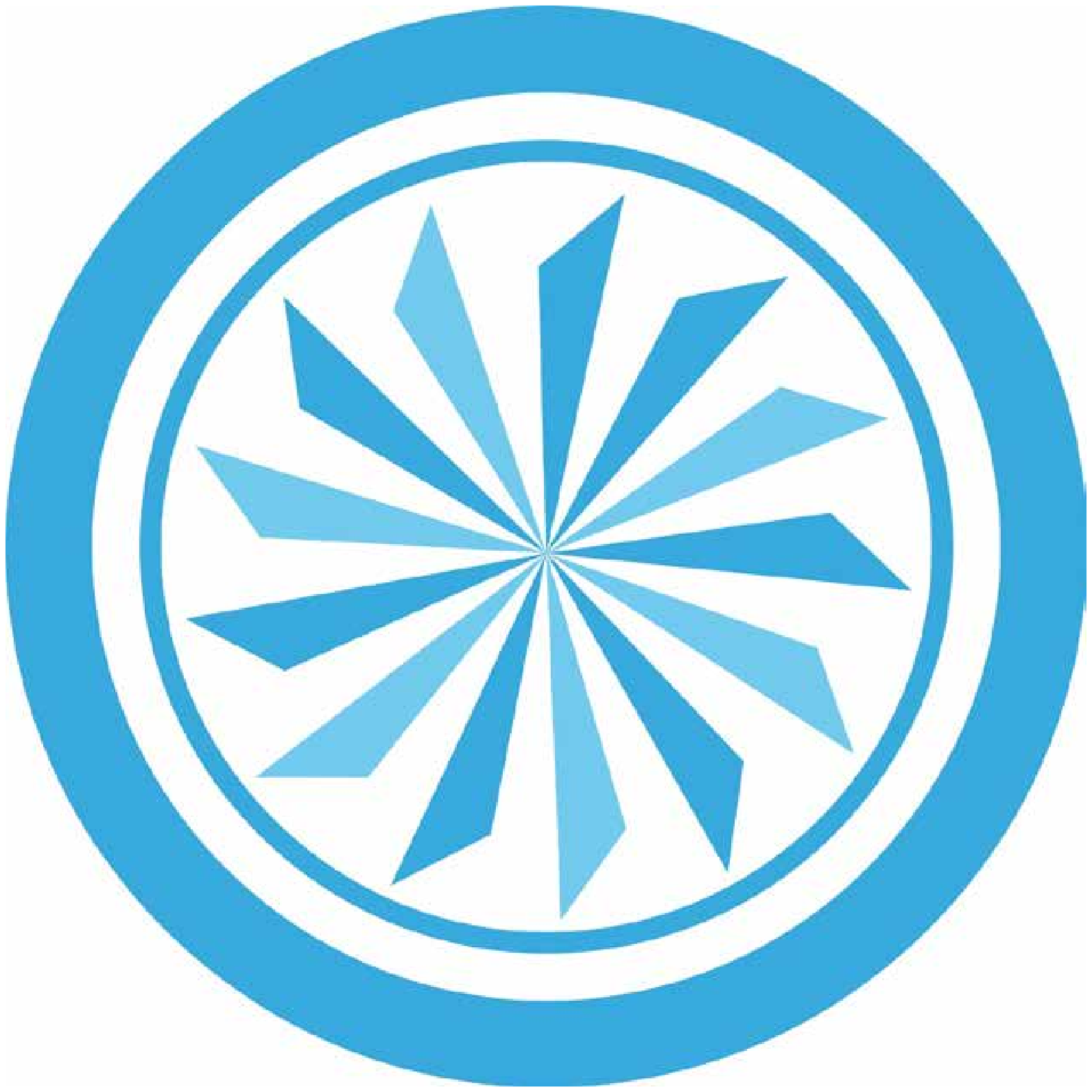}}
~
\begin{center}
\shadowbox{
\begin{minipage}{\textwidth}
 \begin{center}
 \vspace{0.3cm}
 \textbf{\huge{Two-fluids plane or axisymmetric}\\
              \vspace{0.3cm}
              \huge{Poiseuille flow at optimum}}\\
 \vspace{0.3cm}
\huge{A (quasi)analytical study}\\
\vspace{0.3cm}
 \end{center}
 \end{minipage}
 }
\vspace{0.2cm}

 \begin{large}
     {\bf Ivan Fedioun}\\
     Polytech Orléans, 8 rue Léonard de Vinci\\
	   45072 Orléans cedex 2, France\\
		{\small ivan.fedioun@univ-orleans.fr}\\
		 \vspace{0.3cm}
			April 2023
   \end{large}
	\end{center}
}

\begin{small}
The solution of the two-fluids plane or axisymetric Poiseuille flow is derived analytically. Then, the conditions for the maximum flow rate of the most viscous fluid are analyzed in terms of fluids volume fractions. The axisymmetric case is totally analytical, whereas the plane one is not. In the axisymmetric case, it is shown that an optimum can be found only if the most viscous fluid (the liquid) is inside and the less viscous one (the gas) is outside. In the plane case, the maximum flow rate of the "liquid" can reach 4 times that of the classical Poiseuille flow in the limit of vanishing viscosity of the "gas".

\tableofcontents
\end{small}

\section{Context of the study}\label{sec:Contexte}

This work was initiated in the context of a final project by engineering students at Polytech Orléans, devoted to the optimization of the suction function of a medical endoscope. The manufacturer observed that the flow rate of aspirated body fluid has a maximum when an air inlet is present in the circuit. However, if the air inlet is too large, the performance of the device drops. Liquid-gas two-phase flows are very complex \cite{FITREMANN83} and it is not intended here to address the problem in its entirety. We propose a simple analytical study based on the axisymmetric or plane Poiseuille flow model, which allows to explain to some extent the observed phenomenon. The literature about two-phase Poiseuille flows is scarce, and this study is surprisignly quite original. 

\section{Classical Poiseuille flow: reminder}\label{sec:Poicla}

The plane or axisymmetric Poiseuille flow is one of the very few analytical solutions of Navier-Stokes equations, a classical exercise for any student in fluid mechanics \cite{SCHLICH68}. Under the assumptions of a steady and parallel flow of incompressible fluid ($\rho=$cste), the problem to be solved is
\begin{equation}\label{eq:NS-Pois}
\frac{dp}{dx}=\mu \Delta u = K < 0
\end{equation}
\begin{equation}\label{eq:laplacien}
  \Delta u=  \left\{  \begin{array}{ll} 
                                \displaystyle{\frac{d^2 u}{dy^2}}                                                 & \mbox{: plane}   \\
                                \displaystyle{\frac{1}{r}\frac{d}{dr}\left( r\frac{du}{dr} \right)}  & \mbox{: axisymmetric} 
                                \end{array} 
											\right.
\end{equation}
~\vspace{11pt}

\noindent
\begin{minipage}{0.48\textwidth}
\noindent
$\bullet$ {\it In the axisymmetric case} ($0 \leq r \leq R$), the general solution of (\ref{eq:NS-Pois}) is
\begin{equation}
u(r)=\frac{K}{\mu}\left[ \frac{r^2}{4} + A \ln r + B\right]
\end{equation}
The finite velocity on the axis ($r=0$) imposes $A=0$, and the no-slip condition $u(\pm R)=0$ gives $B=-R^2/4$. The solution is then
\begin{equation}\label{eq:SolGene-Poisaxi}
\boxed{u(r)=U\left[ 1-\left(\frac{r}{R}\right)^2\right]}
\end{equation}
\begin{equation}\label{eq:Umax-Poisaxi}
\boxed{U=-\frac{K}{\mu}\frac{R^2}{4} > 0}
\end{equation}
where $U=2 U_{ave}$ is the maximum velocity along the axis of the pipe. The volume flow rate is
\begin{equation}\label{eq:Q-Poisaxi}
Q^{\mbox{\scriptsize{POIS}}}_{\mbox{\scriptsize{axi}}}=\pi R^2 U_{ave} = -2 \pi \frac{K}{\mu}\frac{R^4}{16} 
\end{equation}
\end{minipage}
\hfill
\begin{minipage}{0.48\textwidth}
\noindent
$\bullet$ {\it In the plane case} ($-H \leq y \leq H$), the general solution of (\ref{eq:NS-Pois}) is
\begin{equation}
u(y)=\frac{K}{\mu}\left[ \frac{y^2}{2} + A y + B\right]
\end{equation}
The maximum velocity in the symmetry plane ($y=0$) imposes $A=0$, and the no-slip condition $u(\pm H)=0$ gives $B=-H^2/2$. The solution is then
\begin{equation}\label{eq:SolGene-Poisplan}
\boxed{u(y)=U\left[ 1-\left(\frac{y}{H}\right)^2\right]}
\end{equation}
\begin{equation}\label{eq:Umax-Poisplan}
\boxed{U=-\frac{K}{\mu}\frac{H^2}{2} > 0}
\end{equation}
where $U=\frac{3}{2} U_{ave}$ is the velocity in the symmetry plane. The volume flow rate ({\it per} m) is
\begin{equation}\label{eq:Q-Poisplan}
Q^{\mbox{\scriptsize{POIS}}}_{\mbox{\scriptsize{plane}}}=2 H U_{ave} = -\frac{2}{3} \frac{K}{\mu} H^3 
\end{equation}
\end{minipage}

\section{Axisymmetric two-fluids Poiseuille flow}\label{sec:Poibi_axi}

\subsection{General solution}

\begin{figure}[!h]
 \begin{center}
 \includegraphics[width=0.6\textwidth]{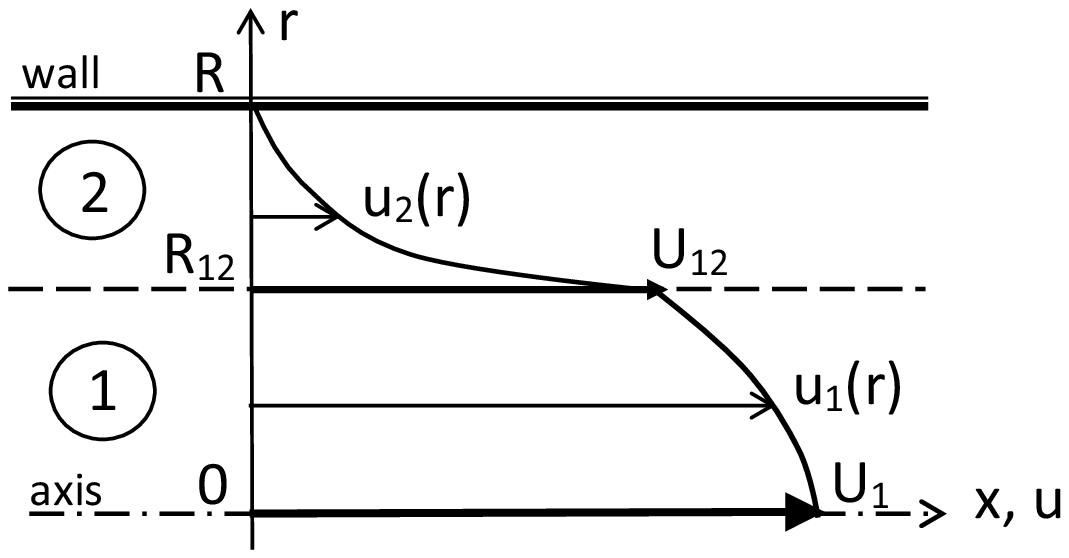}
 \caption{Two-fluids axisymmetric Poiseuille flow configuration.}
 \label{fig:Poibi_axi}
 \end{center}
\end{figure}

Let us consider a two-fluids axisymmetric configuration, fluid 1 being in the center of the pipe, and fluid 2 in an annular configuration around fluid 1 (figure \ref{fig:Poibi_axi}). Fluid 1 is therefore not constrained by the no-slip condition at the wall. The radius $0 \leq R_{12} \leq R$ is at the interface of the two fluids assumed non-miscible. 

In a cross section of the pipe, both fluids are at the same pressure. The problem reads:
\begin{equation}\label{eq:NS-Poisbi_axi}
\frac{dp}{dx}= \frac{\mu_1}{r}\frac{d}{dr}\left( r\frac{du_1}{dr} \right) =  \frac{\mu_2}{r}\frac{d}{dr}\left( r\frac{du_2}{dr} \right) = K < 0
\end{equation}
and the general solution is 
\begin{equation}
u_\alpha(r)=\frac{K}{\mu_\alpha}\left[ \frac{r^2}{4} + A_\alpha \ln r + B_\alpha\right]  \quad ; \quad \alpha= 1, 2
\end{equation}
Boundary conditions are:
\begin{eqnarray}
 r=0  \quad &:&  \frac{du_1}{dr} = 0 \label{CL1_axi} \\
r=R_{12}     &: &  u_1(R_{12})=u_2(R_{12})=U_{12} \label{CL2_axi} \\
                   & & \mu_1 \frac{du_1}{dr} = \mu_2 \frac{du_2}{dr}  \label{CL3_axi} \\
r=\pm R     &:& u_2(\pm R)=0  \label{CL4}
\end{eqnarray} 
We deduce the constants:
\[A_1= A_2 =0 \quad  ; \quad B_1= - \left(\frac{R^2-R_{12}^2}{4 \eta}\right)  - \frac{R_{12}^2}{4} \quad ; \quad B_2= - \frac{R^2}{4}  \]
where
\begin{equation}\label{eq:eta}
\boxed{\eta=\frac{\mu_2}{\mu_1} > 0}
\end{equation}
is the first physical parameter of the problem. If $\eta < 1$, the most viscous fluid (the "liquid") is in the center, side 1, and vice versa. Velocity profiles are:
\begin{equation}\label{eq:u1_axi}
\boxed{u_1(r) = -\frac{K}{\mu_1}\left( \frac{R_{12}^2-r^2}{4} + \frac{R^2-R_{12}^2}{4 \eta}  \right) \quad ; \quad 0 \leq r \leq R_{12}}
\end{equation}
\begin{equation}\label{eq:u2_axi}
\boxed{u_2(r) = -\frac{K}{\mu_2}\left( \frac{R^2-r^2}{4} \right) \quad ; \quad R_{12} \leq r \leq R}
\end{equation}
\vspace{11pt}

The volume flow rates of fluid 1 and fluid 2, functions of $R_{12}$, are obtained from
\begin{equation}\label{eq:Q12_def_axi}
Q_1(R_{12}) = 2 \pi \int_0^{R_{12}} u_1(r)\, r dr  \quad ; \quad Q_2(R_{12}) = 2 \pi \int_{R_{12}}^R u_2(r)\, r dr
\end{equation}
\begin{equation}\label{eq:Q1_axi}
\boxed{Q_1(R_{12}) = -2 \pi \frac{K}{\mu_1}\left[  \frac{R_{12}^4}{16} + \frac{R_{12}^2}{8 \eta} \left(R^2-R_{12}^2\right)\right] }
\end{equation}
\begin{equation}\label{eq:Q2_axi}
\boxed{Q_2(R_{12}) = -2 \pi \frac{K}{\mu_2}\left[ \frac{R^4 +R_{12}^4}{16} - \frac{R^2 R_{12}^2}{8}\right] }
\end{equation}

In the case $\eta=1$, one can check that $Q_1 + Q_2 =Q^{\mbox{\scriptsize{POIS}}}_{\mbox{\scriptsize{axi}}}$ where $Q^{\mbox{\scriptsize{POIS}}}_{\mbox{\scriptsize{axi}}}$ is the flow rate (\ref{eq:Q-Poisaxi}) of the classical axisymmetric Poiseuille flow, and if $R_{12}=R$ (resp. $R_{12}=0$), $Q_1=Q_{1 \mbox{ \scriptsize{axi}}}^{\mbox{\scriptsize{POIS}}}$ (resp. $Q_2=Q_{2 \mbox{ \scriptsize{axi}}}^{\mbox{\scriptsize{POIS}}}$).

\subsection{Search for the optimum}\label{sec:PoibiOptaxi}

The problem is, given the viscosity ratio $\eta$, to find the radius $R_{12}$ - if it exists- that maximizes the flow rate of the most viscous fluid, i.e. the flow rate $Q_1$ if $\eta <1$ or the flow rate $Q_2$ if $\eta > 1$. The flow rate $Q_2$ (\ref{eq:Q2_axi}) does not depend on $\eta$. Therefore, the search for optimum only makes sense for $\eta \leq 1$, i.e. for $Q_1$. Unfortunately, the most physically probable case is $\eta >> 1$: a liquid film on the wall drained by the faster and less viscous gas in the center.\\ 

Let us define:
\vspace{-22pt}
\begin{eqnarray}
                  r &=& R \; \overline{r}  \quad\quad\quad\quad\;\;\;\,  ; \quad 0 \leq \overline{r} \leq 1 \\
        R_{12} &=& R \; \rho  \quad\quad\quad\quad\;\;\;\,  ; \quad 0 \leq \rho \leq 1 \\
   Q_\alpha(R_{12}) &=& Q_\alpha^{\mbox{\scriptsize{POIS}}}  \; \overline{Q}_\alpha(\rho) \quad ; \quad \alpha = 1, 2 \\
u_\alpha(r) &=& U_1 \; \overline{u}_\alpha(\overline{r})
\end{eqnarray}

The reference velocity must be the same for both flows. We choose the maximum velocity $U_1$ (\ref{eq:Umax-Poisaxi}) of the most viscous internal fluid. The flow rates are normalized by those of the respective Poiseuille flows (\ref{eq:Q-Poisaxi}). It gives:
\begin{eqnarray}
\overline{u}_1(\overline{r}) &=& \rho^2-\overline{r}^2 + \frac{1}{\eta}(1-\rho^2)  \quad\quad ; \quad 0 \leq \overline{r} \leq \rho < 1 \label{eq:u1bar_axi} \\
\overline{u}_2(\overline{r}) &=& \left(  1 - \overline{r}^2 \right)/\eta \quad\quad\quad\quad\quad\;\;\, ; \quad 0 < \rho \leq \overline{r} \leq 1 \label{eq:u2bar_axi} 
\end{eqnarray}
\begin{eqnarray}
\overline{Q}_1(\rho) &=&  \frac{2}{\eta} \rho^2 (1 - \rho^2) + \rho^4 \label{eq:Q1bar_axi} \\
\overline{Q}_2(\rho) &=&  1-2 \rho^2 + \rho^4                                    \label{eq:Q2bar_axi} 
\end{eqnarray}

One can see that the second physical, or rather geometrical, parameter on which the problem depends is
\begin{equation}\label{eq:rho}
\boxed{\rho=\frac{R_{12}}{R}}
\end{equation}
from which the volume fractions of both fluids are straightforward:
\begin{equation}\label{eq:X1X2_axi} 
X_1= \rho^2 \quad ; \quad X_2=1 - \rho^2
\end{equation}

\begin{figure}[!h]
\begin{center}
\begin{minipage}{0.48\textwidth}
 \begin{center}
 \includegraphics[width=\textwidth]{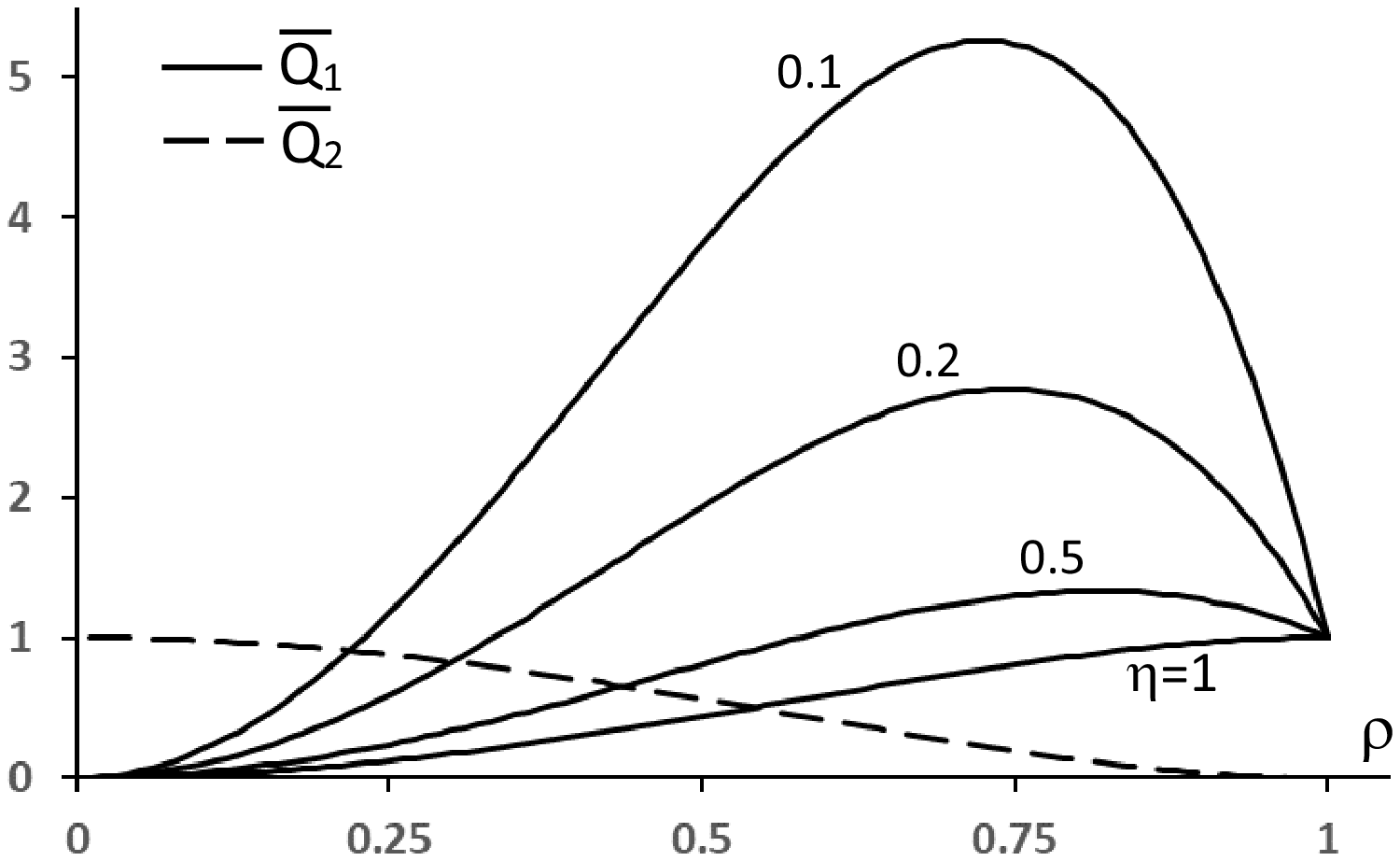}
 \caption{Normalized flow rates $\overline{Q}_1(\rho)$ (\ref{eq:Q1bar_axi}) and $\overline{Q}_2(\rho)$ (\ref{eq:Q2bar_axi}).}
 \label{fig:Q1_axi}
 \end{center}
\end{minipage}
\hfill
\begin{minipage}{0.48\textwidth}
 \begin{center}
 \includegraphics[width=\textwidth]{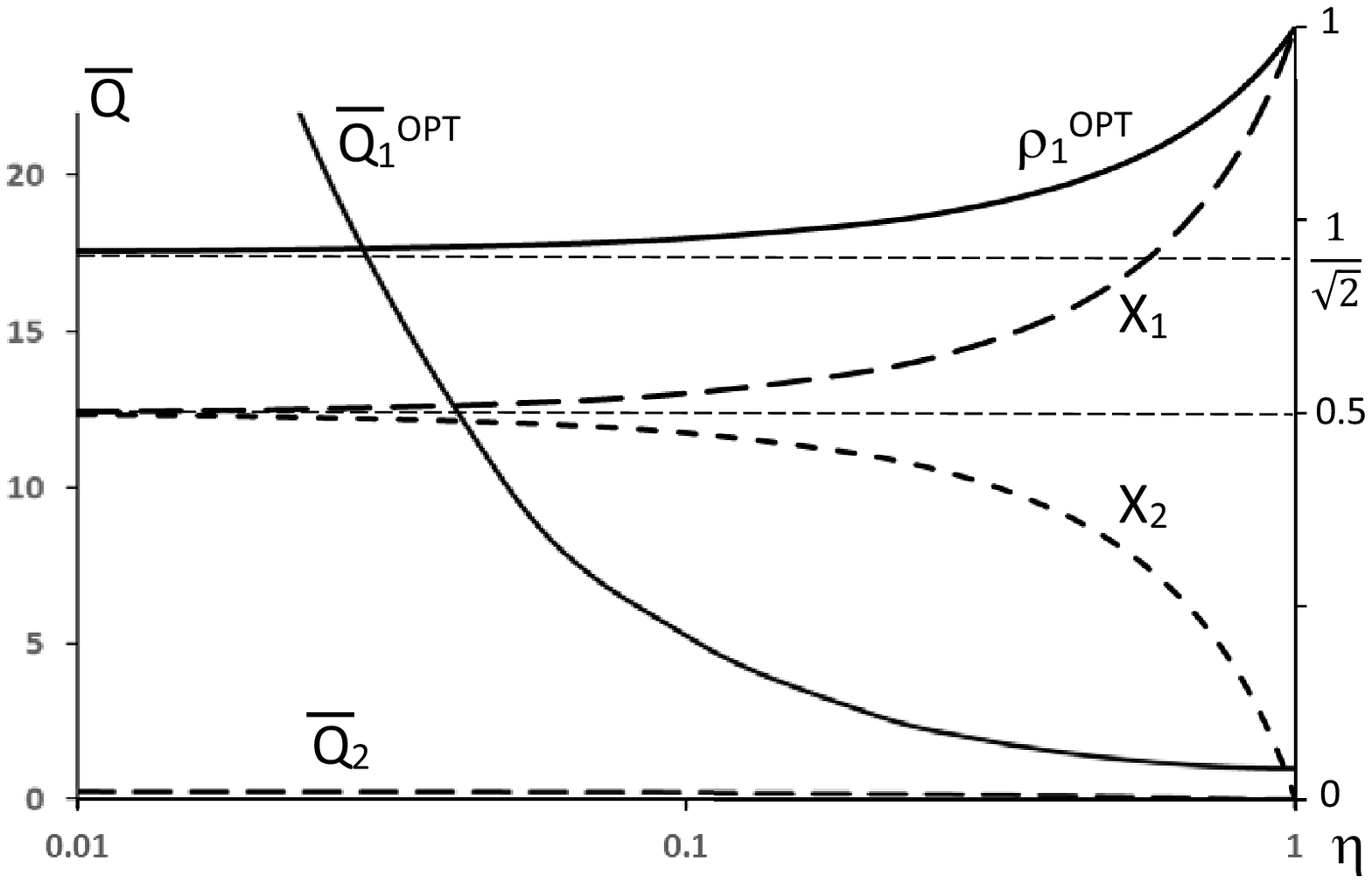}
 \caption{Optimum normalized radius (\ref{eq:Rho1_OPT_axi}), flow rates (\ref{eq:Q1Q2opt}) and volume fractions.}
 \label{fig:Eta_Rho_Q_axi}
 \end{center}
\end{minipage}
\end{center}
\end{figure}

The normalized flow rates $\overline{Q}_1(\rho)$ and $\overline{Q}_2(\rho)$ are displayed in figure \ref{fig:Q1_axi}. Indeed, the flow rate $\overline{Q}_1(\rho)$ shows a maximum that moves from $\rho=1$ to $\rho \approx 0.7$ when $\eta$ decreases. This finding is consistent with the experimental obvervations described in the introduction.

In order to find the maximum flow rate, one writes $\displaystyle{\frac{d \overline{Q}_1}{d \rho} = 0}$, which leads to
\begin{eqnarray}
                       \rho &=& 0 \quad ; \quad\quad\quad\quad\quad\quad\quad\quad\quad\;\, \overline{Q}_1=0 \quad \mbox{trivial minimum} \nonumber\\
1 - (2-\eta) \rho^2 &=& 0 \quad ; \quad  \boxed{\rho{\mbox{\scriptsize{OPT}}} = \frac{1}{\sqrt{2-\eta}}} \quad \overline{Q}_1 \mbox{ maximum sought} \label{eq:Rho1_OPT_axi}
\end{eqnarray}
with limit $1/\sqrt{2} \approx 0.707\cdots$ when $\eta \rightarrow 0$. The flow rates at optimum are then
\begin{equation}\label{eq:Q1Q2opt} 
\boxed{\overline{Q}_1^{\mbox{\scriptsize{OPT}}} =\frac{1}{\eta (2-\eta)}} \quad ; \quad \boxed{\overline{Q}_2 = \left(\frac{1-\eta}{2-\eta}\right)^2}
\end{equation}

It can be seen that a viscosity ratio of 4 ($\eta \leq 0.25$) more than doubles the flow rate of the inner most viscous fluid. All these results are analytical.

\subsection{Velocity profiles}

\begin{figure}[!h]
 \begin{center}
 \includegraphics[width=0.95\textwidth]{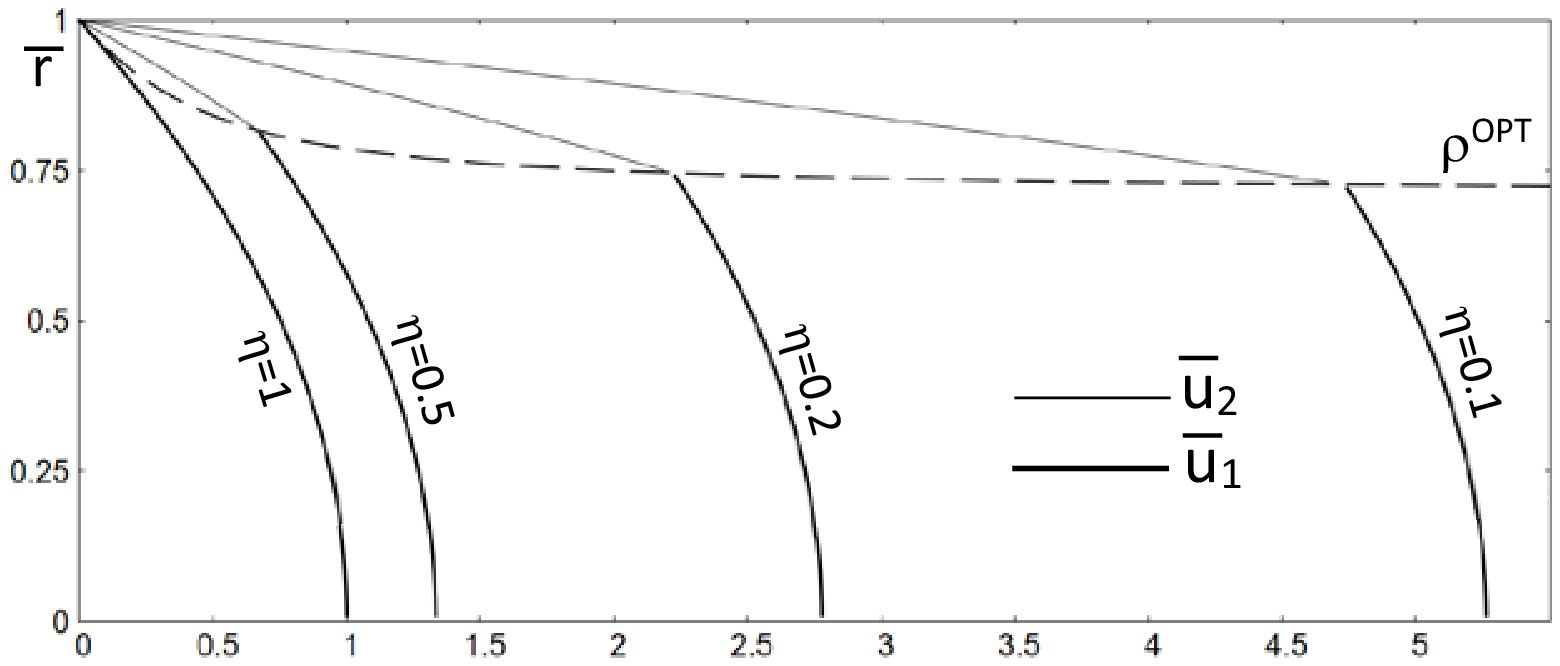} 
 \caption{Velocity profiles that maximize the volume flow rate $Q_1$ ($\eta < 1$).}
 \label{fig:u1_u2_eta_low_axi}
 \end{center}
\end{figure}

Velocity profiles are shown in figure \ref{fig:u1_u2_eta_low_axi}. One can see that the velocity of the "liquid" can reach much higher values than that of the classical Poiseuille flow, by effect of entrainment by the annular gas flow. The maximum velocity along the axis is 
\[\overline{u}_1(0)  =\frac{1}{\eta (2-\eta)}=\overline{Q}_1^{\mbox{\scriptsize{OPT}}}\]
and can theoretically reach infinity in the limit $\eta \rightarrow 0$.\\

\section{Plane two-fluids Poiseuille flow}\label{sec:Poibi_plan}

\subsection{General solution}

\begin{figure}[!h]
 \begin{center}
 \includegraphics[width=0.6\textwidth]{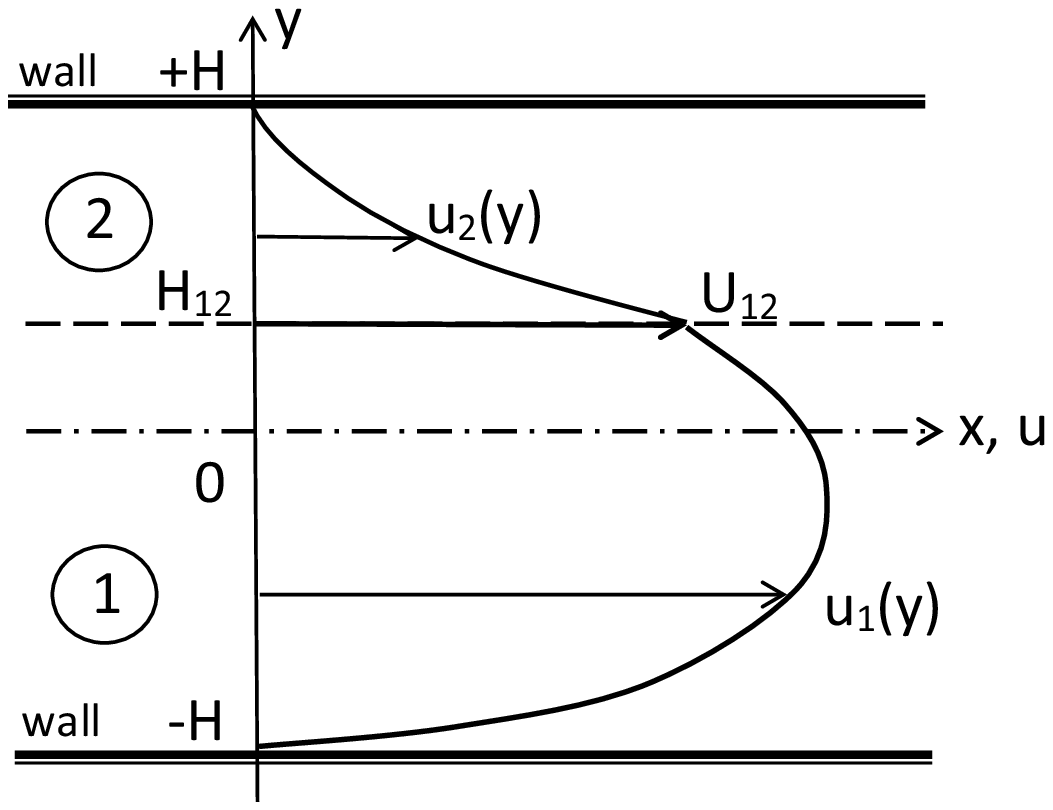}
 \caption{Two-fluids plane Poiseuille flow configuration.}
 \label{fig:Poibi_plan}
 \end{center}
\end{figure}

Let us now consider a plane two-fluids configuration, with fluid 1 at the bottom of the channel and fluid 2 at the top (figure \ref{fig:Poibi_plan}), with height $-H \leq H_{12} \leq +H$ splitting the non-miscible two fluids. Unlike the axisymmetric case, both fluids are now constrained by the no-slip condition at the walls. Hence, there exists a symmetry {\it top $\leftrightarrow$ bottom} for viscosity ratios $\eta \leftrightarrow (1/\eta)$. If $\eta < 1$, the most viscous fluid is at the bottom and vice versa.

The problem reads:
\begin{equation}\label{eq:NS-Poisbi_plan}
\frac{dp}{dx}= \mu_1 \frac{d^2u_1}{dy^2} =  \mu_2 \frac{d^2u_2}{dy^2} = K < 0
\end{equation}
and the general solution is 
\begin{equation}
u_\alpha(y)=\frac{K}{\mu_\alpha}\left[ \frac{y^2}{2} + A_\alpha y + B_\alpha\right]  \quad ; \quad \alpha= 1, 2
\end{equation}
Boundary conditions are:
\begin{eqnarray}
y= -H            &:&  u_1(-H) = 0 \label{CL1_plan} \\
y=H_{12}      &: &  u_1(H_{12})=u_2(H_{12})=U_{12} \label{CL2_plan} \\
                     & & \mu_1 \frac{du_1}{dy} = \mu_2 \frac{du_2}{dy}  \label{CL3_plan} \\
y=+H             &:& u_2(+H)=0  \label{CL4_plan}
\end{eqnarray} 
We deduce the constants:
\begin{equation}\label{eq:A}
A_1= A_2 = \frac{A}{2} = \frac{(\eta-1) (H^2 - H_{12}^2)}{2 \left[ (\eta+1)H + (\eta -1)H_{12} \right]}
\end{equation}
\[B_1= \frac{H}{2} (A - H) \quad ; \quad B_2= - \frac{H}{2} (A + H)  \]

Velocity profiles, matching the solution given by Vincent {\it et al.} \cite{VCL04}, are:
\begin{equation}\label{eq:u1_plan}
\boxed{u_1(y) = -\frac{K}{\mu_1}\left[ \frac{H^2-y^2}{2} - \frac{A}{2} (H + y) \right] \quad ; \quad -H \leq y \leq H_{12}}
\end{equation}
\begin{equation}\label{eq:u2_plan}
\boxed{u_2(y) = -\frac{K}{\mu_2}\left[ \frac{H^2-y^2}{2} + \frac{A}{2} (H - y) \right] \quad ; \quad H_{12} \leq y \leq H}
\end{equation}
\vspace{11pt}

The volume flow rates of fluid 1 and fluid 2, functions of $H_{12}$, are obtained from
\begin{equation}\label{eq:Q12_def_plan}
Q_1(H_{12}) = \int_{-H}^{H_{12}} u_1(y) dy  \quad ; \quad Q_2(H_{12}) = \int_{H_{12}}^H u_2(y) dy
\end{equation}
\begin{equation}\label{eq:Q1_plan}
\boxed{Q_1(H_{12}) = -\frac{K}{\mu_1}\left[ \frac{ H^2 (H+H_{12})}{2} - \frac{ H^3 +H_{12}^3}{6} - \frac{A}{2} \left( H ( H +H_{12}) - \frac{ H^2 -H_{12}^2}{2}\right)\right] }
\end{equation}
\begin{equation}\label{eq:Q2_plan}
\boxed{Q_2(H_{12}) = -\frac{K}{\mu_2}\left[ \frac{ H^2 (H+H_{12})}{2} - \frac{ H^3 -H_{12}^3}{6} + \frac{A}{2} \left( H ( H -H_{12}) - \frac{ H^2 -H_{12}^2}{2}\right)\right] }
\end{equation}

\subsection{Search for the optimum}\label{sec:PoibiOptplan}

Again, given a value $0 < \eta \leq 1 $, we want to find the height $H_{12}$ that maximizes the flow rate $Q_1$ of the most viscous fluid (the problem is symmetric).\\ 

Let us define:
\vspace{-22pt}
\begin{eqnarray}
                  y &=& H \; \overline{y}  \quad\quad\quad\quad\;\;\;\,  ; \quad -1 \leq \overline{y} \leq 1 \\
        H_{12} &=& H \; \rho  \quad\quad\quad\quad\;\;\;\,  ; \quad -1 \leq \rho \leq 1 \\
   Q_\alpha(H_{12}) &=& Q_\alpha^{\mbox{\scriptsize{POIS}}}  \; \overline{Q}_\alpha(\rho) \quad ; \quad \alpha = 1, 2 \\
u_\alpha(y) &=& U_1 \; \overline{u}_\alpha(\overline{y})
\end{eqnarray}

The reference velocity is the maximum velocity $U_1$ (\ref{eq:Umax-Poisplan}) of the most viscous fluid. Flow rates are normalized by those of the respective Poiseuille flows (\ref{eq:Q-Poisplan}). It gives:
\begin{eqnarray}
\overline{u}_1(\overline{y}) &=&   \; 1-\overline{y}^2 -  \overline{A} ( 1 + \overline{y})   \quad\quad\quad ; \quad -1 \leq \overline{y} \leq \rho < 1 \label{eq:u1bar_plan} \\
\overline{u}_2(\overline{r}) &=& \left[ 1-\overline{y}^2 +  \overline{A} ( 1 - \overline{y})  \right]/\eta \quad\; ; \quad -1 < \rho \leq \overline{y} \leq 1 \label{eq:u2bar_plan} 
\end{eqnarray}
\begin{eqnarray}
\overline{Q}_1(\rho) &=&  \frac{3}{4} \left[ 1 +\rho - \frac{1}{3} (1 + \rho^3) - \frac{\overline{A}}{2} (1 +\rho)^2 \right]  \label{eq:Q1bar_plan} \\
\overline{Q}_2(\rho) &=&  \frac{3}{4} \left[ 1 - \rho - \frac{1}{3} (1 - \rho^3) + \frac{\overline{A}}{2} (1 - \rho)^2 \right]  \label{eq:Q2bar_plan} 
\end{eqnarray}
with
\begin{equation}\label{eq:Abar}
 \overline{A}(\rho)= \frac{(\eta-1) (1 - \rho^2)}{\eta+1 + (\eta -1)\rho}
\end{equation}
The volume fractions of both fluids are
\begin{equation}\label{eq:X1X2_plan} 
X_1= \frac{1}{2} (1+\rho) \quad ; \quad X_2=\frac{1}{2} (1-\rho)
\end{equation}

\begin{figure}[!h]
\begin{center}
\begin{minipage}{0.47\textwidth}
 \begin{center}
 \includegraphics[width=\textwidth]{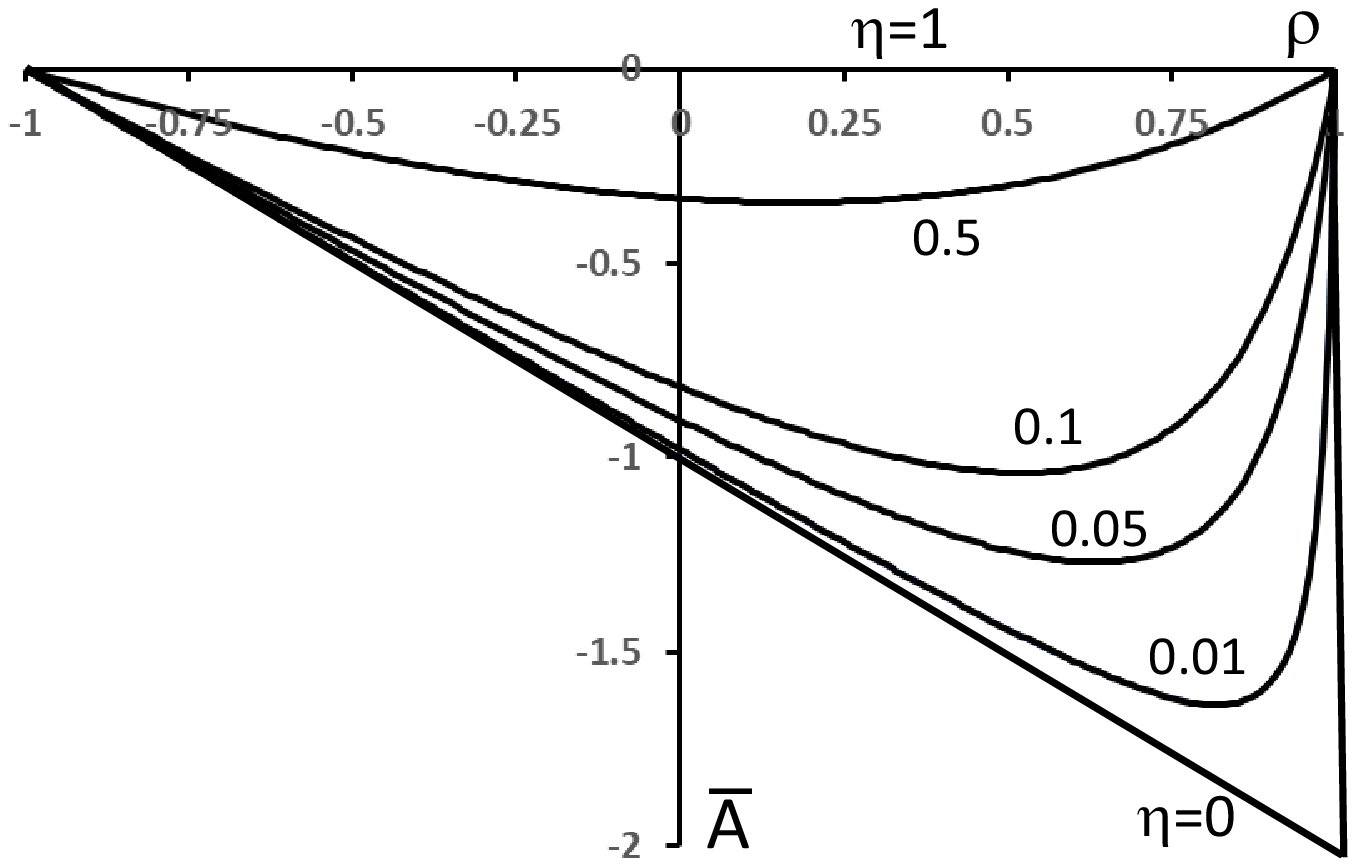}
 \caption{Function $\overline{A}$, eq.(\ref{eq:Abar}).}
 \label{fig:Aplan}
 \end{center}
\end{minipage}
\hfill
\begin{minipage}{0.49\textwidth}
 \begin{center}
 \includegraphics[width=\textwidth]{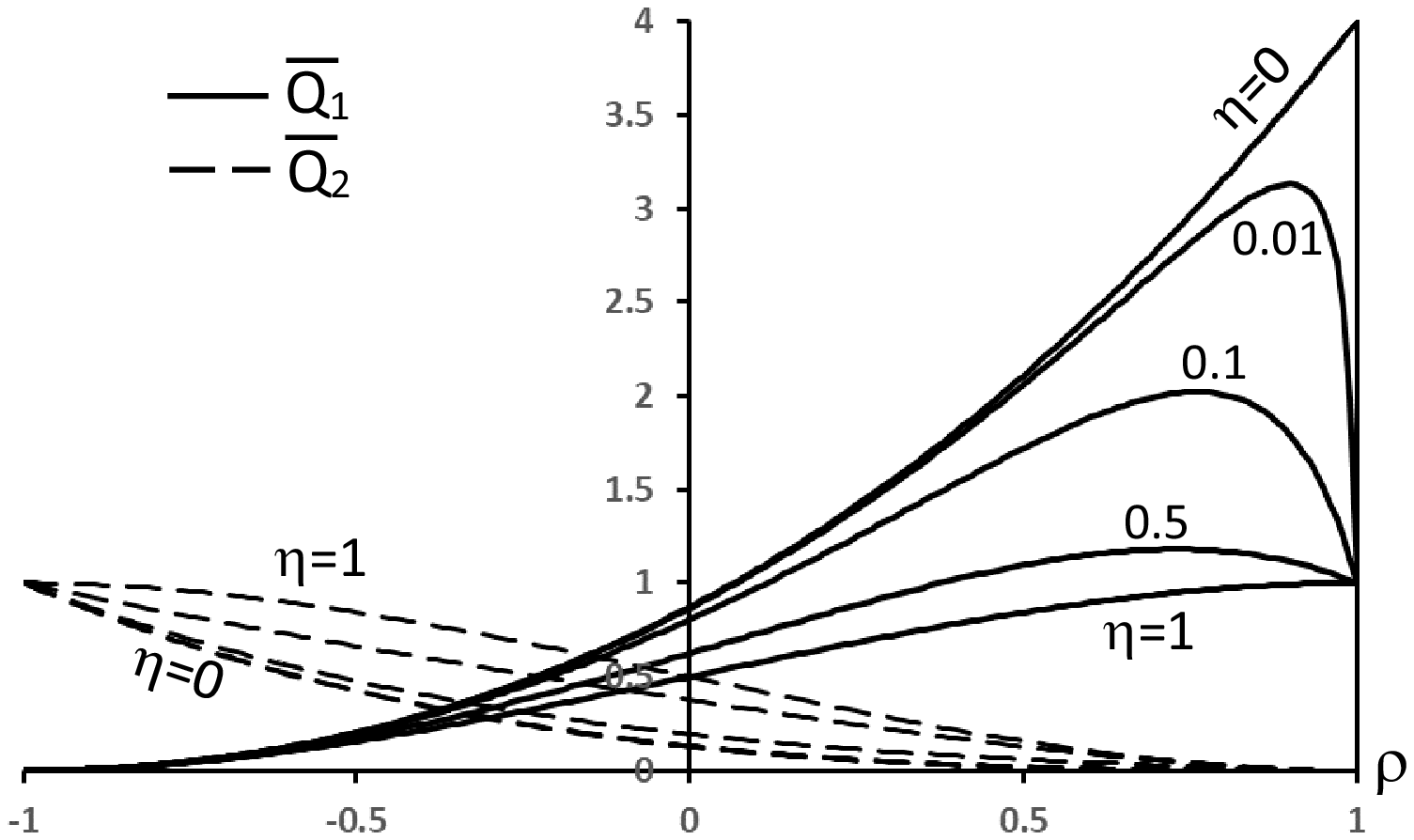}
 \caption{Flow rates $\overline{Q}_1(\rho)$ and $\overline{Q}_2(\rho)$.}
 \label{fig:Q1_plan}
 \end{center}
\end{minipage}
\end{center}
\end{figure}

The function $\overline{A}(\rho)$ with parameter $\eta$, from which the solutions differ compared to the classical Poiseuille flow, is shown in figure \ref{fig:Aplan}. Its limit when $\eta \rightarrow 0$ is $\overline{A}_{\eta \rightarrow 0}(\rho)=-(1+\rho)$, for $-1 \leq \rho < 1$ and $\overline{A}_{\eta \rightarrow 0}(1)=0$. The normalized flow rates $\overline{Q}_1(\rho)$ and $\overline{Q}_2(\rho)$ are displayed in figure \ref{fig:Q1_plan}. Flow rate $\overline{Q}_1(\rho)$ shows a maximum that moves as $\eta$ decreases, but non-monotonically unlike the axisymmetric case (figures \ref{fig:Q1_axi} and \ref{fig:Eta_Rho_Q_axi}). 

The optimum flow rate 1 is obtained from $\displaystyle{\frac{d \overline{Q}_1}{d \rho} = 0}$, that leads to
\begin{eqnarray}
                                                                                                                  \rho = -1 \quad &;& \quad \overline{Q}_1=0 \quad \mbox{trivial minimum} \nonumber\\
\boxed{1 - \rho -\overline{A}-\frac{\overline{A}'}{2} (1+ \rho) = 0  = f_1(\rho)}  \quad &;& \quad \overline{Q}_1 \mbox{ maximum sought} \label{eq:Rho1_OPT_plan}
\end{eqnarray}
where $\overline{A}'(\rho)$ is the derivative of (\ref{eq:Abar}) with respect to $\rho$. Symmetrically, we get for flow rate 2:
\begin{eqnarray}
                                                                                                                  \rho = +1 \quad &;& \quad \overline{Q}_2=0 \quad \mbox{trivial minimum} \nonumber\\
\boxed{1 + \rho + \overline{A}-\frac{\overline{A}'}{2} (1- \rho) = 0 = f_2(\rho) } \quad &;& \quad \overline{Q}_2 \mbox{ maximum sought} \label{eq:Rho2_OPT_plan}
\end{eqnarray}

\begin{figure}[!h]
\begin{center}
\begin{minipage}{0.48\textwidth}
 \begin{center}
 \includegraphics[width=\textwidth]{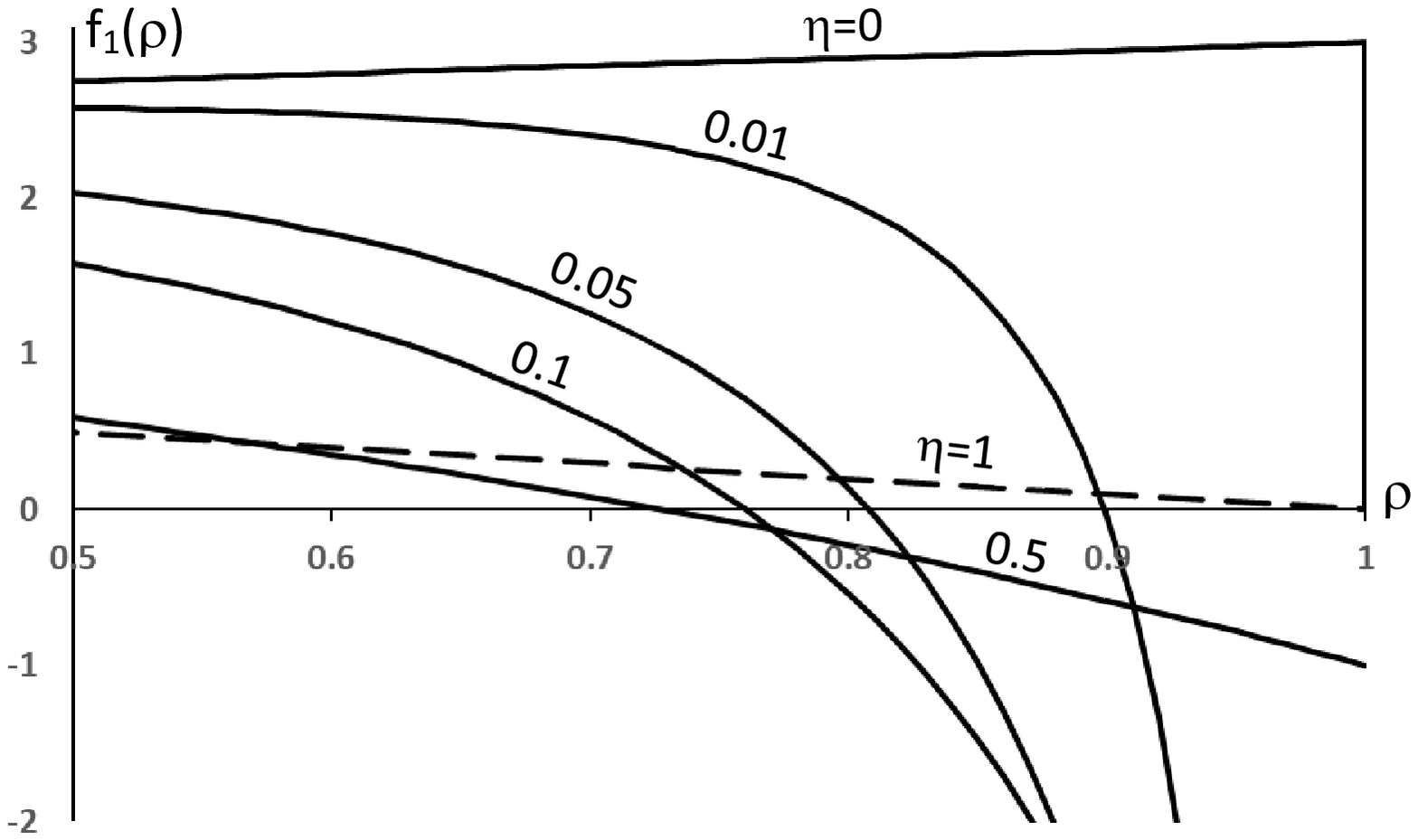}
 \caption{Optimization function  (\ref{eq:Rho1_OPT_plan}).}
 \label{fig:f1_plan}
 \end{center}
\end{minipage}
\hfill
\begin{minipage}{0.48\textwidth}
 \begin{center}
 \includegraphics[width=0.95\textwidth]{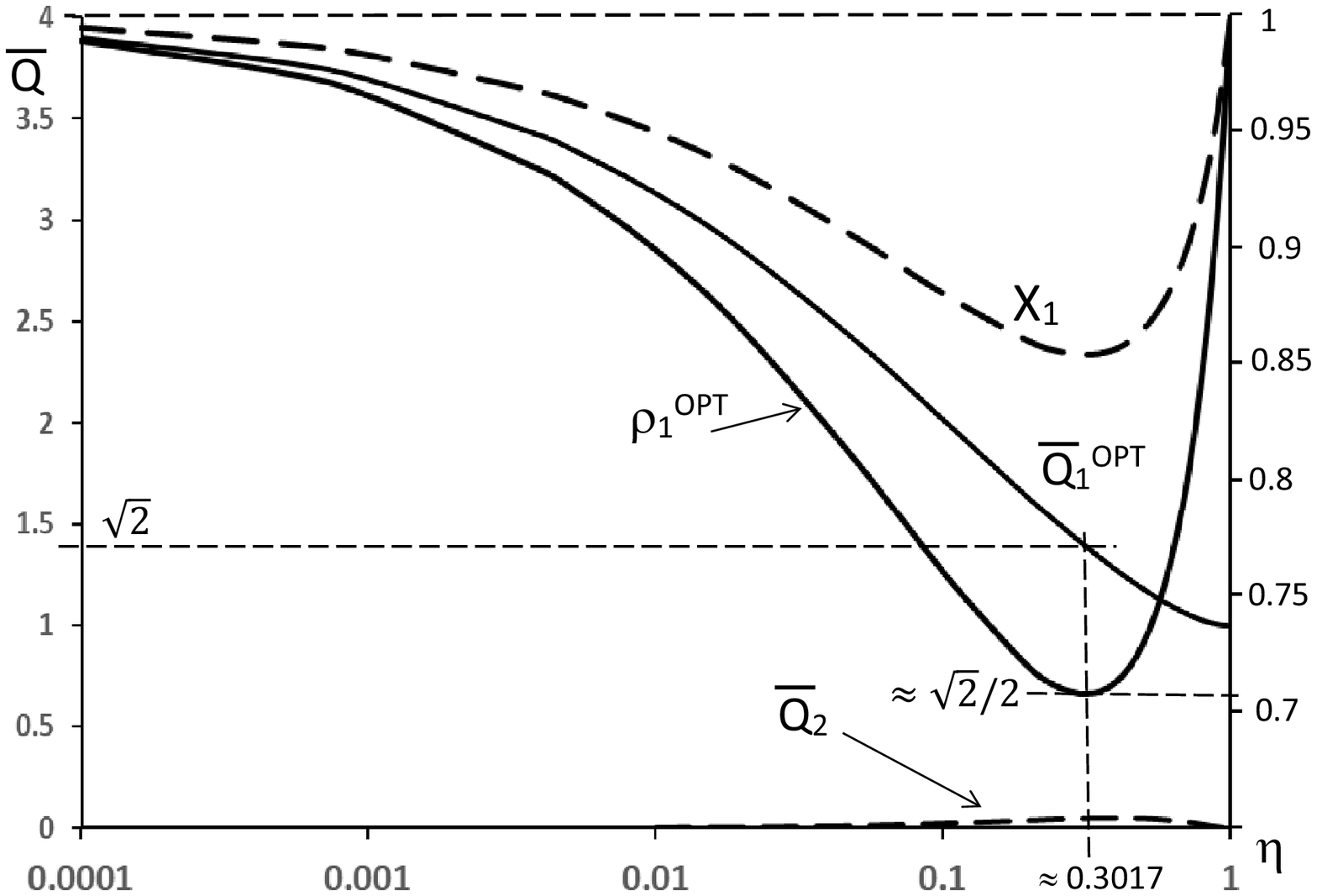}
 \caption{Optimum normalized height, volume flow rates and volume fractions.}
 \label{fig:Eta_Rho_Q_plan}
 \end{center}
\end{minipage}
\end{center}
\end{figure}

The optimization function $f_1(\rho)$ is shown in figure \ref{fig:f1_plan} for different values of $\eta$. Its roots $\rho_1^{\mbox{\scriptsize{OPT}}}(\eta)$ are not analytical. They are computed with the Newton-Raphson method (absolute convergence at $10^{-12}$). They are plotted in figure \ref{fig:Eta_Rho_Q_plan} and some numerical values are given in table \ref{tab:Eta_Rho} in appendix.

It is remarkable in figure \ref{fig:Eta_Rho_Q_plan} that at the specific point where $\rho_1^{\mbox{\scriptsize{OPT}}}$ is minimum, the optimum flow rate is $\overline{Q}_1^{\mbox{\scriptsize{OPT}}} \approx \sqrt{2}$ and $\rho_1^{\mbox{\scriptsize{OPT}}} \approx \sqrt{2}/2$ . If we accept these values, we can get after some algebra the theoretical corresponding viscosity ratio $\eta$ and flow rate $\overline{Q}_2$:
\[ \eta = \frac{1}{5}\left(\frac{47 -24\sqrt{2}}{3+4\sqrt{2}}\right) \approx 0.30170023 \quad ; \quad \overline{Q}_2 = \frac{3}{2}\left(\frac{3 -2\sqrt{2}}{3+2\sqrt{2}}\right) \approx 0.04415588 \]
 
With this theoretical value of $\eta$, the Newton-Raphson method converges to $\rho_1^{\mbox{\scriptsize{OPT}}} = 0.70721664\cdots \approx \sqrt{2}/2$ within $1.1\, 10^{-4}$ (-0.015\%),  $\overline{Q}_1^{\mbox{\scriptsize{OPT}}} \approx \sqrt{2}$ within $3.8\, 10^{-8}$ (-0.0000027\%), and $\overline{Q}_2 = 0.044128\cdots$, which is again very close to the theoretical value. It was not possible to further investigate this issue, neither analytically nor numerically, but this specific point reminds us the asymptotic limit $\sqrt{2}/2$ of the optimal radius (\ref{eq:Rho1_OPT_axi}) of the axisymmetric case. Asymptotic behaviors of different variables for limit values of $\eta$ are gathered in table \ref{tab:Eta_vers_0}.
\begin{table}[!h]
\begin{center}
\begin{tabular}{|c|c|c|c|}
\hline
$\eta$                                                               & 0 \quad ($-1 \leq \rho < 1$)    & 1		           &  $+\infty$ \quad ($-1 < \rho \leq 1$) \\
\hline
$A$                                                                    &  $-(1+\rho)$                              & 0              &  $1-\rho$   \\
$A'$                                                                   &  $-1$                                          & 0              &  $-1$  \\
$f_1(\rho)$                                                       & $2+ (1 +\rho)/2$                       & $1-\rho$ & $(1+\rho)/2$ \\
$f_2(\rho)$                                                       & $(1-\rho)/2$                              & $1+\rho$ &  $2+ (1 -\rho)/2$ \\
$\rho_1^{\mbox{\scriptsize{OPT}}}$               & 1                                                &  1             &    --- \\
$\overline{Q}_1^{\mbox{\scriptsize{OPT}}}$  & 4                                                &  1              &   --- \\
\hline
\end{tabular}
\caption{Limit behaviors of solutions.}
\label{tab:Eta_vers_0}
\end{center}
\end{table}
~\vspace{-22pt}

\subsection{Velocity profiles}
\begin{figure}[!h]
 \begin{center}
 \includegraphics[width=0.9\textwidth]{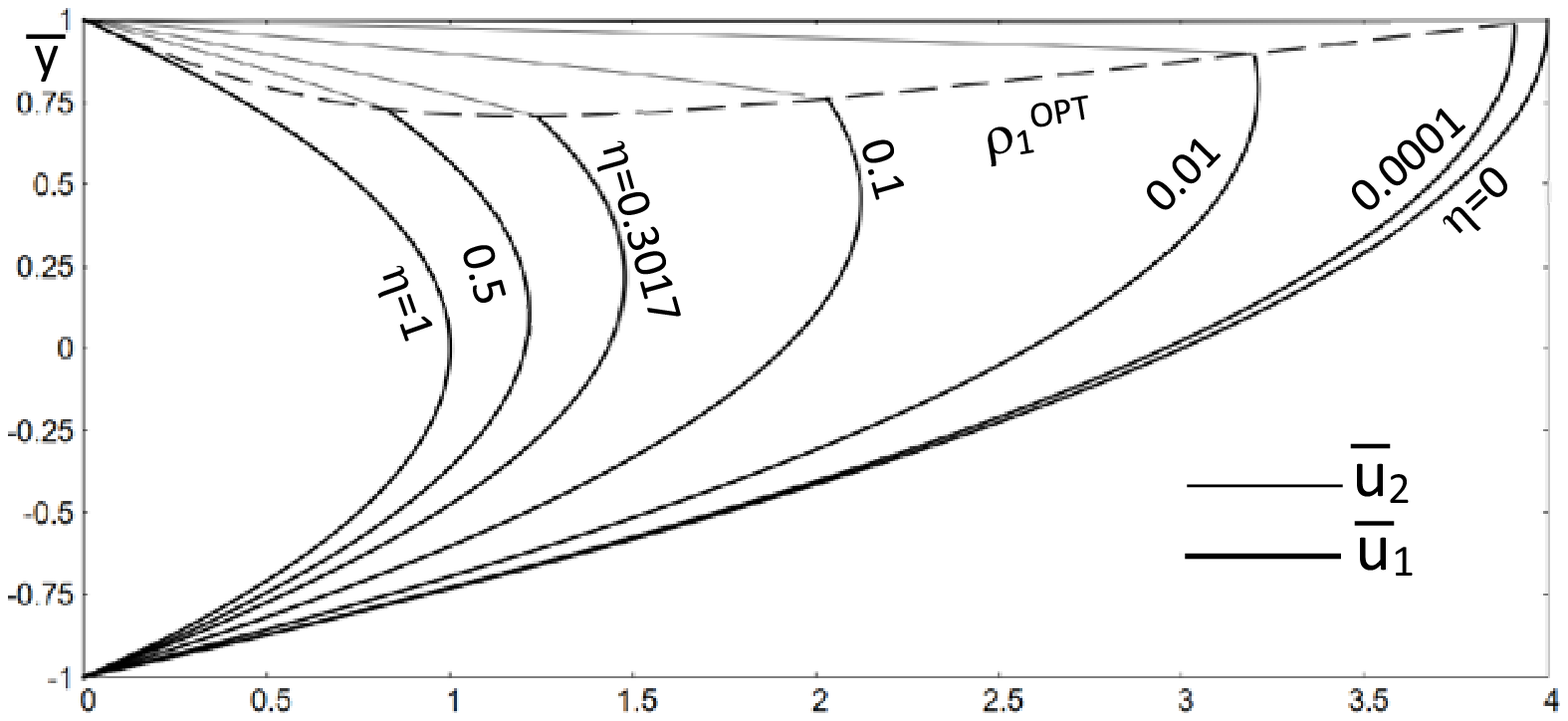} 
 \caption{Velocity profiles that maximize the volume flow rate $Q_1$ ($\eta < 1$).}
 \label{fig:u1_u2_eta_plan}
 \end{center}
\end{figure}

Velocity profiles are shown in figure \ref{fig:u1_u2_eta_plan}. As in the axisymmetric case, one can see that the "liquid", drained by the less viscous gas, reaches higher velocities than that of the classical Poiseuille flow. The maximum velocity is obtained for $\eta \rightarrow 0$:  
\[\overline{u}_{1, \eta \rightarrow 0}(\overline{y})  =(1+\overline{y})(3-\overline{y}) \quad \mbox{with} \quad \overline{u}_{1, \eta \rightarrow 0}(1)  = 4 = \overline{Q}_{1, \eta \rightarrow 0}^{\mbox{\scriptsize{OPT}}}  \]
~\vspace{-5pt}
~\vspace{-11pt}

\section{Conclusions and perspectives}\label{sec:concl}

This analytical study is of course very schematic and does not pretend to represent all the complexity of two-phase flows in pipes. However, it has allowed us to identify a mechanism that can explain the experimental observations. To complete it, one could introduce the surface tension forces at the liquid-gas interface.

The axisymmetric case is unfortunately optimal for $\eta <1$ only (liquid in the center, gas around). Optimization of the $\eta >1$ case (gas in the center, liquid at the walls) would have been more interesting for industrial applications. However, the flow rate of the central liquid can be theoretically infinite in the limit $\eta \rightarrow 0$.

The plane case is no more analytical. The maximum reachable flow rate is 4 times that of the classical Poiseuille flow in the limit $\eta \rightarrow 0$. The specific point $\rho_1^{\mbox{\scriptsize{OPT}}}=\sqrt{2}/2$, $\overline{Q}_1^{\mbox{\scriptsize{OPT}}}=\sqrt{2}$ is striking, but is not an exact optimum.\\

The two-phase "slug" type of flow, where long bubbles of gas are drained at the center of the pipe, then collapse and split again, could be modeled alternating $\eta >1 \leftrightarrow \eta < 1$ with an intermittency factor to be designed more or less empirically, in order to reproduce the experimental optimum. 
 The question arises of course as how to achieve this optimum for a given viscosity ratio $\eta$, i.e. how to concretely realize the air intake in the pipe (parietal, axial...) and how to calibrate its flow rate, if not by the practitioner's feeling.   

\begin{figure}[!ht]
 \begin{center}
 \includegraphics[width=\textwidth]{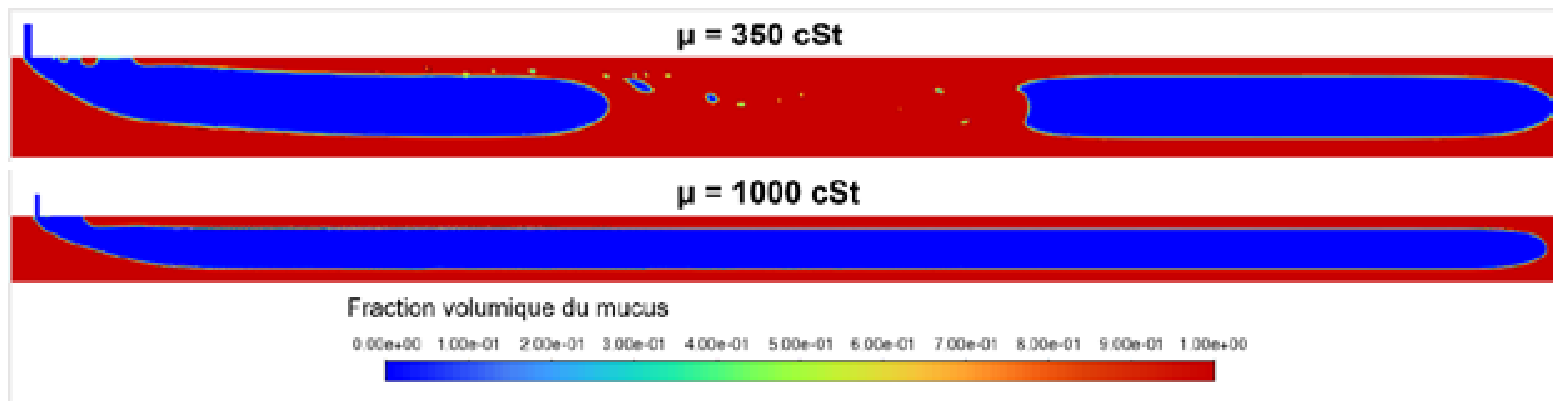}
 \caption{VOF ANSYS/Fluent$^\circledR$ simulation of a plane (2D) "slug" flow (Delavault A. / Maberi-Grodet M., student's project, Polytech Orléans 2023)}
 \label{fig:slug}
 \end{center}
\end{figure}

\bibliography{Biblio_Pois}

\clearpage

\section*{Appendix}

\begin{table}[!h]
\begin{center}
\begin{tabular}{|c|c|c|c|c|c|}
\hline
$\eta$ &	$\rho_1^{\mbox{\scriptsize{OPT}}}$ &	$\overline{Q}_1^{\mbox{\scriptsize{OPT}}}$	& $\overline{Q}_2^{\mbox{\scriptsize{OPT}}}$	&$X_1$	&$X_2$ \\
\hline
   0     &    1                 &          4          &           0           &            1         &          0          \\
 0.0001 & 0.988607239 & 3.898045090 & 0.000001856 & 0.99430362 & 0.00569638 \\
 0.0005 & 0.974943430 & 3.777271089 &	0.000019820 & 0.987471714 & 0.012528286\\
   0.001 & 0.988607239 & 3.898045090 & 0.000054198 & 0.994303619 & 0.005696380 \\
  0.005  & 0.925693963 & 3.354808198 & 0.000526320 & 0.962846981 & 0.037153019 \\
    0.01  & 0.898894644 & 3.132626433 & 0.001341430 & 0.949447322 & 0.050552678 \\
     0.02 & 0.864583746 & 2.854801700 & 0.003280031 & 0.932291873 & 0.067708127 \\
     0.05 & 0.807820625 & 2.405606500 & 0.009748029 & 0.903910312 & 0.096089688 \\
       0.1 & 0.759984915 & 2.021779694 & 0.020006793 & 0.879992457 & 0.120007543 \\
       0.2 & 0.718661678 & 1.631679570 & 0.035428058 & 0.859330839 & 0.140669161 \\
     0.25 & 0.710431818 & 1.511426731 & 0.040530850 & 0.855215909 & 0.144784091 \\
       0.3 & 0.707250263 & 1.417070415 & 0.044035615 & 0.853625132 & 0.146374868 \\
			{\bf 0.30170023}& {\bf 0.707216642} &	{\bf 1.414213601} & {\bf 0.044128111} &	{\bf 0.853608321} &	{\bf 0.146391679} \\
     0.35 & 0.708055811 & 1.340840836 & 0.046059677 & 0.854027905 & 0.145972095 \\
       0.4 & 0.712221090 & 1.278007399 & 0.046714672 & 0.856110545 & 0.143889455 \\
     0.45 & 0.719368769 & 1.225491404 & 0.046106818 & 0.859684385 & 0.140315615 \\
    0.5 & 0.729281773 & 1.181191183 & 0.044340018 & 0.864640887 & 0.135359113 \\
  0.55 & 0.741856425 & 1.143621618 & 0.041520381 & 0.870928213 & 0.129071787 \\
    0.6 & 0.757077754 & 1.111707824 & 0.037761984 & 0.878538877 & 0.121461123 \\
  0.65 & 0.775007768 & 1.084660856 & 0.033194269 & 0.887503884 & 0.112496116 \\
    0.7 & 0.795782520 & 1.061900134 & 0.027971887 & 0.897891260 & 0.102108740 \\
  0.75 & 0.819616365 & 1.043004186 & 0.022288332 & 0.909808182 & 0.090191818 \\
    0.8 & 0.846813471 & 1.027679729 & 0.016395568 & 0.923406736 & 0.076593264 \\
  0.85 & 0.877788146 & 1.015743714 & 0.010633270 & 0.938894073 & 0.061105927 \\
    0.9 & 0.913097411 & 1.007115719 & 0.005473940 & 0.956548705 & 0.043451295 \\
  0.95 & 0.953492312 & 1.001820172 & 0.001595138 & 0.976746156 & 0.023253844 \\
       1 & 1 & 1 & 0 & 1 & 0 \\
\hline
\end{tabular}
\caption{Numerical roots of (\ref{eq:Rho1_OPT_plan}), and corresponding optimum solutions. In bold, numerical solution at the remarkable point, very close to $\rho_1^{\mbox{\scriptsize{OPT}}}=\sqrt{2}/2$, $\overline{Q}_1^{\mbox{\scriptsize{OPT}}}=\sqrt{2}$.}
\label{tab:Eta_Rho}
\end{center}
\end{table}

\end{document}